# Surface waves and surface stability for a pre-stretched, unconstrained, non-linearly elastic half-space

## J.G. Murphy

*Department of M echanical Engineering, Dublin City University, Dublin 9, IRELAND.*

## Michel Destrade

*School of Electrical, Electronic, and Mechanical Engineering, University College Dublin, Belfield, Dublin 4, IRELAND.*

**Abstract**

An unconstrained, non-linearly elastic, semi-infinite solid is maintained in a state of large static plane strain. A power-law relation between the pre-stretches is assumed and it is shown that this assumption is well-motivated physically and is likely to describe the state of pre-stretch for a wide class of materials. A general class of strain-energy functions consistent with this assumption is derived. For this class of materials, the secular equation for incremental surface waves and the bifurcation condition for surface instability are shown to reduce to an equation involving only ordinary derivatives of the strain-energy equation. A compressible neo-Hookean material is considered as an example and it is found that finite compressibility has little quantitative effect on the speed of a surface wave and on the critical ratio of compression for surface instability.

## 1. Introduction

Since first considered by Rayleigh [1], the problem of waves propagating along the surface of an elastic half-space has proved to be of relevance to seismologists and engineers alike. One particular aspect of this problem that is still being investigated is that of the effect of a pre-stress on the speed of propagating waves and on stability analysis. The pioneering works of Hayes and Rivlin [2] and of Biot [3] considered this problem in the setting of an incremental perturbation superimposed upon the large static deformation of a pre-stressed, homogeneous, isotropic, non-linearly elastic half-space. These authors assume plane strain and consider perturbations with sinusoidal variations along one principal direction in the plane boundary and exponential decay along the principal direction normal to the boundary. Most of the work since has been within this context.

There have been essentially two approaches: the first assumes that the solid is perfectly incompressible whereas the second assumes the material is unconstrained, although recovery of the incompressibility theory in some limiting process is usually considered. Half-spaces of the incompressible Mooney material were considered by Biot [3], Flavin [4], and Willson [5]. Willson [6, 7] also studied the propagation of surface waves for general incompressible half-spaces. The approach of Dowaikh and Ogden [8] is independent of the choice of strain invariants for general incompressible half-spaces and has proved very influential. See, for example, Rogerson and Fu [9], Rogerson and Sandiford [10], or Destrade and Scott [11].

The unconstrained approach initiated by Hayes and Rivlin [2] was further developed by Chadwick and Jarvis [12] and by Dowaikh and Ogden [13]. The latter paper gives a mathematical description of surface waves propagating along a principal direction of the underlying pure homogeneous strain in a general compressible half-space and is the starting point of this study. The novelty here is that a power-law relation between the pre-stretches is assumed *ab initio* for plane strain deformations. Not only is this assumption mathematically tractable, enabling a corresponding strain-energy function to be derived in a natural way, but also, this relation is compatible with experimental data for a large class of non-linearly elastic materials, both solids and foams. See Section 3 for a discussion.

In Section 4 the corresponding wave-speed secular equation and bifurcation criterion are shown to depend on the strain-energy function only through *ordinary* derivatives of the strain-energy function with respect to one of the pre-stretches. This situation is analogous to the structure of the wave speed equation and of the bifurcation criterion for perfectly incompressible elastic materials, as given, for example, in Dowaikh and Ogden [8]. It is further shown that as the compressibility goes to zero, the perfectly incompressible equation is recovered as a special case. Surprisingly, it is shown that the compressibility has little effect on the relations between pre-stretch and wave speed or critical compression stretch ratio for a particular example, which can be thought of as a compressible neo-Hookean material.

The ultimate aim of the paper is to quantify how finite compressibility affects surface stability and surface wave propagation in pre-strained solids. Usually ideal incompressibility is often assumed simply because it simplifies the mathematical analysis, by removing one material parameter. For instance, it is known [14] that all Mooney materials (characterized by two material constants) present surface instability in plane strain at the critical compression stretch ratio of 0.544 whereas their compressible counterpart, the Hadamard materials, have a critical compression stretch ratio which depends markedly on the material parameters. It has been experimentally observed [15] that the surface of compressed rubber blocks, often modelled as incompressible Mooney solids, buckles much earlier than predicted by theory, and it is thus natural to wonder if slight compressibility could provide an explanation for this disparity. As this article demonstrates, the effect of compressibility on the stability of neo-Hookean solids (belonging to the Mooney class) is negligible. Similar conclusions are reached about surface wave propagation.

## 2. The boundary condition

For half-spaces occupied by a homogeneous, isotropic, hyperelastic material, consider the plane, pure homogeneous deformation of the form

$$x = \lambda X, \ y = \gamma Y, \ z = Z, \tag{1}$$

where *(X,Y,Z)* and *(x,y,z)* denote the Cartesian coordinates of a particle before and after deformation respectively, and $\lambda, \gamma$ are constants. It is assumed that the deformed solid occupies the region $y < 0$ with boundary $y = 0$. Denoting the corresponding principal stretches by $\lambda_i, i = 1, 2, 3$, the following identification is made in this case: $(\lambda_1, \lambda_2, \lambda_3) = (\lambda, \gamma, 1)$.

For ease of exposition, it is assumed here that the boundary of the half-space is free of tractions, although a non-vanishing normal load on the boundary can be easily accommodated in an obvious way. Therefore for a given strain-energy function *W* it is required that

$$W_2(\lambda_1, \lambda_2, 1) = 0, \tag{2}$$

where the subscript notation attached to *W* denotes partial differentiation with respect to the appropriate principal stretch, evaluated at $\lambda_3 = 1$. Then, if $W_{22} \neq 0$, there exists a function $f(\lambda_1)$ such that

$$\lambda_2 = f(\lambda_1). \tag{3}$$

In practice, however, it is difficult, if not impossible, to obtain this relation explicitly in terms of elementary functions, which obviously makes computation and analysis difficult. As an illustration of the difficulties encountered, consider the restricted Hadamard material characterized by the strain-energy function

$$W = \frac{\mu}{2}\left(\lambda_1^2 + \lambda_2^2 + \lambda_3^2 - 3\right) + \kappa H(i_3), \ i_3 = \lambda_1 \lambda_2 \lambda_3, \tag{4}$$

where $\mu, \kappa$ are respectively the (positive) infinitesimal shear and bulk moduli and the function *H* is restricted such that

$$H(1) = 0, H'(1) = -\mu/\kappa, H''(1) = 1 + (\mu/3\kappa), \tag{5}$$

where the prime notation denotes ordinary differentiation. This material has received much attention in the literature. For example, this material was studied within the current context by Chadwick and Jarvis [12] and by Dowaikh and Ogden [13], within the context of surface instabilities by Usmani and Beatty [16] and within the context of finite wave propagation by Boulanger, Hayes and Trimarco [17], Boulanger and Hayes [18] and Rodrigues-Feirrera *et al.* [19]. For this material

$$\frac{\partial W}{\partial \lambda_2} = \mu \lambda_2 + \kappa H'(i_3)\lambda_1 \lambda_3, \quad \frac{\partial^2 W}{\partial \lambda_2^2} = \mu + \kappa H''(i_3)\lambda_1^2 \lambda_3^2, \tag{6}$$

from which it follows that if $H$ is a convex function, then $W_{22} > 0$, which is assumed henceforth. For the traction-free boundary condition (2), equation (6)$_1$ yields

$$\lambda_2^2 = K(i_3), \ K(i_3) \equiv -(\kappa/\mu)i_3 H'(i_3), \tag{7}$$

and therefore

$$\lambda_2 = f(\lambda_1), \ f^{-1}(x) \equiv x^{-1}K^{-1}(x^2). \tag{8}$$

This equation indicates that, with the exception of especially simple forms for the arbitrary function $H(i_3)$, determination of the function $f$ in (3) is not possible. In practice, as can be seen in Chadwick and Jarvis [12] and in Dowaikh and Ogden [13], this means that what is a one-dimensional problem in terms of $\lambda_1$, say, is analysed as if it were a two-dimensional problem in both $\lambda_1, \lambda_2$. Since determination of the relation (3) is essential for the full analysis of Rayleigh waves, an inverse approach is adopted here: assuming a *physically well-motivated form* for *f*, the stress-free boundary condition is integrated to obtain a general class of materials that is compatible with this form.

**3. Modelling the relation between the stretches**

Assume that the principal stretches are related through a *power-law expression* of the form

$$\lambda_2 = \lambda_1^{\varepsilon - 1}, \ 0 < \varepsilon < 1. \tag{9}$$

The motivation for choosing this relation is not mere mathematical convenience: there is a strong physical motivation for this form as well. This seemingly artificial relation is, in fact, widely applicable for non-linearly elastic materials, of which there are two main types: solid and foamed rubbers. The classical assumption made when modelling solid rubbers is that they are perfectly incompressible. However this is an idealization because all solid rubbers are to some extent compressible; it is proposed here that this compressibility effect for pure homogeneous, plane strain deformations of solid rubbers is captured through (9), where it is likely that

$$0 < \varepsilon \ll 1, \tag{10}$$

(because setting $\varepsilon = 0$ in (9) yields $\lambda_2 = \lambda_1^{-1}$, the relation predicted by perfect incompressibility). The main experimental motivation for the power-law assumption (9), (10) for *solid rubbers* is given by Beatty and Stalnaker [20], who obtained experimental data for both the stretch in the direction of the applied force, denoted here by $\lambda_1$, and the two equal stretches in the perpendicular directions $\lambda_2$ and $\lambda_3 = \lambda_2$, for six different rubbers in *simple tension*. The following power-law kinematic relationship is shown to match the data extremely well:

$$\lambda_2 = \lambda_1^{\varepsilon - (1/2)}, \tag{11}$$

where $\varepsilon$ is a *small* (relative to one), positive material parameter (for example, for a urethane sample used by Beatty and Stalnaker $\varepsilon = 0.007$). Further evidence to support (11) in simple tension was obtained for the solid rubber NR70 by Bechir et al. [21], who show that (11) yields an almost perfect fit with their experimental data with $\varepsilon = 0.02$. The authors are unaware of any data for plane strain, pure homogeneous deformations of solid rubbers where both stretches are measured. However it seems reasonable to infer that the power-law form that works so well for simple tension can also model deformations of the form (1) for solid rubbers.

Power-law relations between the stretches for pure homogeneous deformations are very common also for *foamed rubbers*. For a foamed polyurethane rubber, Blatz and Ko [22] show that a power-law relation between the two stretches gives an excellent fit for the three material characterisation tests of simple tension, biaxial tension, and strip-biaxial tension. For example in simple tension, it is shown that an excellent fit with data is given by (11) but with $\varepsilon = 0.25$. In strip-biaxial tension one of the three principal stretches is held identically equal to one, corresponding very closely with the deformation considered here. Storåkers [23] performed the same three experiments on two vulcanized foam rubbers, one a natural and the other a synthetic rubber. In each case, he showed that a power-law relation between the stretches gave an excellent fit with the experimental data. Consequently, it is reasonable to assume that, for half-spaces of many foamed rubbers subjected to a deformation of the form (1), a relation of the form (9) holds with small, but finite values, for $\varepsilon$.

## 4. Determination of the strain-energy function

A general class of materials consistent with both (2) and (9) is now derived. Since plane-strain deformations are being considered here, the problem under study is inherently two-dimensional. Let

$$W^*(\lambda_1, \lambda_2) = W(\lambda_1, \lambda_2, 1). \tag{12}$$

Then, for the materials considered here, if $i = 1,2$,

$$W_i = \left.\frac{\partial W(\lambda_1, \lambda_2, \lambda_3)}{\partial \lambda_i}\right|_{\lambda_3 = 1} = \frac{\partial W^*(\lambda_1, \lambda_2)}{\partial \lambda_i}, W_{ii} = \left.\frac{\partial^2 W(\lambda_1, \lambda_2, \lambda_3)}{\partial \lambda_i^2}\right|_{\lambda_3 = 1} = \frac{\partial^2 W^*(\lambda_1, \lambda_2)}{\partial \lambda_i^2}. \tag{13}$$

Now let $W_{SF}(\lambda_1, \lambda_2, \lambda_3)$ denote the set of strain-energy functions for which the stress-free boundary condition (2) yields the relation (9) between the principal stretches. Let

$$W^*_{SF}(\lambda_1, \lambda_2) = W_{SF}(\lambda_1, \lambda_2, 1). \tag{14}$$

The stress-free boundary condition for such materials can therefore be written as

$$W^*_{SF,2}(\lambda_1, \lambda_2) = 0, \tag{15}$$

where the comma notation attached to $W^*_{SF}$ denotes partial differentiation with respect to the appropriate principal stretch. Differentiating (15) with respect to $\lambda_1$ then yields

$$W^*_{SF,12} + (\varepsilon - 1)\lambda_1^{\varepsilon-2} W^*_{SF,22} = 0. \tag{16}$$

Now let

$$\tilde{W}(\lambda_1) \equiv W^*_{SF}(\lambda_1, \lambda_1^{\varepsilon-1}, 1). \tag{17}$$

Then, using the prime notation to denote ordinary differentiation with respect to $\lambda_1$ and noting (15), it follows that

$$\tilde{W}' = W^*_{SF,1}, \tag{18}$$

and

$$\tilde{W}'' = W^*_{SF,11} + (\varepsilon - 1)\lambda_1^{\varepsilon-2} W^*_{SF,12}, \tag{19}$$

or alternatively, using (16),

$$\tilde{W}'' = W^*_{SF,11} - (\varepsilon - 1)^2 \lambda_1^{2(\varepsilon-2)} W^*_{SF,22}. \tag{20}$$

It is shown next that an infinity of strain-energy functions exists that are consistent with both (2) and (9). For homogeneous, isotropic, non-linearly elastic materials in plane strain, the strain-energy function can be written as an arbitrary function of any two, independent, symmetric functions of the principal stretches. For example, assume that $W = W^*(I, J)$, where

$$I = \lambda_1^2 + \lambda_2^2, \quad J = \lambda_1 \lambda_2. \tag{21}$$

The traction-free boundary condition (2) can therefore be written in the form

$$2\frac{\partial W^*}{\partial I}\lambda_2 + \frac{\partial W^*}{\partial J}\lambda_1 = 0, \tag{22}$$

which, using (9), can be expressed in terms of $\lambda_1$ alone as

$$2\frac{\partial W^*}{\partial I} + \frac{\partial W^*}{\partial J}\lambda_1^{2-\varepsilon} = 0. \tag{23}$$

It follows immediately from (9) again that

$$\lambda_1 = J^{1/\varepsilon}, \tag{24}$$

and therefore (23) can be re-written as

$$2\frac{\partial W^*}{\partial I} + \frac{\partial W^*}{\partial J}J^{\frac{2-\varepsilon}{\varepsilon}} = 0. \tag{25}$$

Assuming now that (25) holds for *all* plane strain deformations means that (25) becomes a first-order, linear partial differential equation in $W^*$ which can be easily integrated to yield

$$W^* = A(x), x \equiv I - 2 + \frac{\varepsilon}{1-\varepsilon}\left(J^{\frac{2(\varepsilon-1)}{\varepsilon}} - 1\right), \tag{26}$$

where $A$ is an arbitrary function. Conversely, noting that

$$\frac{\partial W^*}{\partial \lambda_2} = 2A'(x)\left(\lambda_2 - J^{\frac{\varepsilon-2}{\varepsilon}}\lambda_1\right), \tag{27}$$

it follows that the relation (9) is recovered. Thus $W^*$ is consistent with both (2) and (9). It follows immediately that an infinity of such strain-energy functions exists since $I$ could be chosen as, for example, $I = \lambda_1^n + \lambda_2^n, n \neq 0$. In what follows, $W^*$ will be used as an example of the family of strain-energy functions $W_{SF}^*$, defined above. Note that $W^*$ can be written as a function of the stretch $\lambda_1$ only, because

$$x = \lambda_1^2 + \lambda_1^{2(\varepsilon-1)} - 2 + \frac{\varepsilon}{1-\varepsilon}\left(\lambda_1^{2(\varepsilon-1)} - 1\right). \tag{28}$$

To ensure zero strain-energy in the reference configuration, it is required that $A(0) = 0$. From (27), it immediately follows that the stress is identically zero in the refer-ence configuration and that, although $\partial \hat{W}/\partial \lambda_2$ is identically zero for plane strain deformations for which (9) holds, it is not identically zero in general, even in plane strain. To ensure that the classical, linear form of the strain-energy function is recovered on restriction to infinitesimal deformations, it is required that

$$\mu = 2A'(0), \kappa = \frac{4(3-2\varepsilon)}{\varepsilon} A'(0). \tag{29}$$

Motivated by the first of these equations, it is additionally assumed that

$$dA/dx > 0, \tag{30}$$

for all allowable values of *x*.

It follows immediately from (29) that

$$\frac{\mu}{\kappa} = \frac{3\varepsilon}{2(3-2\varepsilon)}, \tag{31}$$

or alternatively, that the infinitesimal Poisson ratio ν is given by

$$\nu = \frac{1-\varepsilon}{2-\varepsilon}. \tag{32}$$

Note that for infinitesimal values of $\varepsilon$, (31) reduces to

$$\frac{\mu}{\kappa} \approx \frac{\varepsilon}{2} + H.O.T. \, in \, \varepsilon. \tag{33}$$

Thus the kinematic assumption (9), with $\varepsilon$ small, leads to the restriction

$$\mu/\kappa \ll 1, \tag{34}$$

which is usually imposed *ab initio* for solid rubbers that are assumed to be almost incompressible.

Noting (29)$_1$, linearising the general strain-energy function (26) yields

$$W^* = \frac{\mu}{2}\left(I - 2 + \frac{\varepsilon}{1-\varepsilon}\left(J^{\frac{2(\varepsilon-1)}{\varepsilon}} - 1\right)\right), \tag{35}$$

which is a plane strain, reduced Hadamard material (see (4)). This strain-energy function is the two-dimensional form of the strain-energy function obtained by Blatz and Ko [22], who, although basing their own method on a power-law relationship between the stretches in simple tension, used a different derivation. It is also the two-dimensional form of the strain-energy function introduced independently by Levinson and Burgess [24] for infinitesimal values of $\varepsilon$. For such infinitesimal values of $\varepsilon$, this strain-energy can be viewed as a natural generalization of the *in*compressible neo-Hookean form

$$W = \frac{\mu}{2}(I - 2). \tag{36}$$

It is also the plane-strain form of the particular reduced Hadamard material introduced by Usmani and Beatty [16], who assumed that $H(i_3) = \alpha(i_3^n - 1)$, constant $\alpha, n$.

## 5. Equations of incremental motion

Following Hayes and Rivlin [2], an infinitesimal motion is now superimposed on the deformation (1) such that the particle with position $(x, y, z)$ is displaced to a point $(\bar{x}, \bar{y}, \bar{z})$ where

$$\bar{x} = x + \delta u(x, y, t),\ \bar{y} = y + \delta v(x, y, t),\ \bar{z} = z, \tag{37}$$

where $\delta$ is a constant such that its squares and higher powers can be neglected in comparison with unity, and $u$, $v$ are the components of the incremental mechanical displacement. Time-harmonic surface waves propagating along the $x$ principal direction of the following form are considered:

$$u = A\exp(sky + i\omega t - ikx),\ v = B\exp(sky + i\omega t - ikx), \tag{38}$$

where $A$, $B$ are constants, $\omega$ is the frequency and $k$ is the wave number. This is the form used by Dowaikh and Ogden [13]. Also in (38), $s$ is the attenuation factor, which makes the perturbation decay with distance from the boundary when $\mathrm{Re}(s) > 0$. The wave speed $c$ is therefore given by $c = \omega/k$, while an incremental *static* deformation is described by (38) at $\omega = 0$.

Substitution of (37), (38) into the equations of motion and satisfaction of the boundary conditions (see, for example, Dowaikh and Ogden [13]) yields the following *secular equation* for the speed $c$ of surface waves for zero normal stress on the boundary of a pre-stretched, homogeneous, compressible, isotropic, elastic half-space:

$$(\gamma_2 \alpha_{22})^{1/2}(\alpha_{11} - \rho_r c^2)^{1/2}(\gamma_1 - \gamma_2 - \rho_r c^2) + (\gamma_1 - \rho_r c^2)^{1/2}(\alpha_{11}\alpha_{22} - \alpha_{12}^2 - \alpha_{22}\rho_r c^2) = 0, \tag{39}$$

where $\rho_r$ is the mass density of the material in the reference configuration and, in the notation of Dowaikh and Ogden [13],

$$\alpha_{11} \equiv \lambda_1^2 W_{11},\ \alpha_{22} \equiv \lambda_2^2 W_{22},\ \alpha_{12} \equiv \lambda_1 \lambda_2 W_{12},$$

$$\gamma_1 \equiv \frac{\lambda_1^2(\lambda_1 W_1 - \lambda_2 W_2)}{\lambda_1^2 - \lambda_2^2}, \ \gamma_2 \equiv \frac{\lambda_2^2(\lambda_2 W_2 - \lambda_1 W_1)}{\lambda_2^2 - \lambda_1^2} = \frac{\lambda_2^2}{\lambda_1^2}\gamma_1. \tag{40}$$

Equation (39) is a re-arranged form of equation (5.11) of Dowaikh and Ogden [13]. Using an obvious squaring process and setting $\lambda_1 = 1$ yields

$$X^3(\breve{\lambda}+2) - 8X^2(\breve{\lambda}+2) + 8X(3\breve{\lambda}+4) - 16(\breve{\lambda}+1) = 0, \ X \equiv \rho_r c^2/\mu, \ \breve{\lambda} \equiv \lambda/\mu, \tag{41}$$

which is precisely the Rayleigh cubic of the linear theory [1].

The wave speed equation (39) holds for all compressible strain-energy functions and therefore holds in particular for the class of materials of interest here, $W_{SF}(\lambda_1,\lambda_2,1)$, defined in the previous section. It follows immediately from (13) that the partial derivatives of $W_{SF}$ can be replaced by the corresponding partial derivatives of $W_{SF}^*$ and therefore (16) yields

$$\alpha_{12} = (1-\varepsilon)\alpha_{22}. \tag{42}$$

Noting also that

$$\gamma_1 = \frac{\lambda_1^3 W_{SF,1}^*}{\lambda_1^2 - \lambda_1^{2(\varepsilon-1)}}, \ \gamma_2 = \lambda_1^{2(\varepsilon-2)}\gamma_1, \tag{43}$$

then (39) yields the following form of the secular equation in terms of partial derivatives of $W_{SF}^*$:

$$\left(\frac{\lambda_1 W_{SF,1}^*}{(\lambda_1^2 - \lambda_1^{2(\varepsilon-1)})W_{SF,22}^*}\right)^{1/2}(\lambda_1^2 W_{SF,11}^* - \rho_r c^2)^{1/2}(\lambda_1 W_{SF,1}^* - \rho_r c^2) +$$
$$\left(\frac{\lambda_1^3 W_{SF,1}^*}{\lambda_1^2 - \lambda_1^{2(\varepsilon-1)}} - \rho_r c^2\right)^{1/2}\left(\lambda_1^2(W_{SF,11}^* - (\varepsilon-1)^2 \lambda_1^{2(\varepsilon-2)} W_{SF,22}^*) - \rho_r c^2\right) = 0, \tag{44}$$

which gives $c$ as an implicit function of $\lambda_1$ only. It must be additionally assumed here that

$$\rho_r c^2 \leq \min\left(\frac{\lambda_1^3 W_{SF,1}^*}{\lambda_1^2 - \lambda_1^{2(\varepsilon-1)}}, \lambda_1^2 W_{SF,11}^*\right), \tag{45}$$

which, on restriction to infinitesimal deformations, yields

$$\rho_r c^2 \leq \min(\mu, \lambda + 2\mu). \tag{46}$$

If this last condition is satisfied, then the Rayleigh equation has a unique, positive solution, as demonstrated, for example, in Atkin and Fox [25]. Attention is therefore focused here on the range of $\lambda_1$ values for which (44) also yields positive solutions. By continuity, this range is bounded by those values of $\lambda_1$ for which $c = 0$ in (44). These $\lambda_1$ values are called the *critical stretches* and are given by the *bifurcation equation*

$$\left(\frac{W^*_{SF,11}}{W^*_{SF,22}}\right)^{1/2} W^*_{SF,1} + \lambda_1\left(W^*_{SF,11} - (\varepsilon-1)^2 \lambda_1^{2(\varepsilon-2)} W^*_{SF,22}\right) = 0, \tag{47}$$

assuming that $W^*_{SF,11} > 0$.

## 6. The critical stretches: an example

To make further progress with either the wave speed equation or the critical stretch equation, the strain-energy function must be specified. Setting $W^*_{SF}(\lambda_1, \lambda_2) = A(x)$, the general strain function given in (26), yields

$$W^*_{SF,22} = \frac{4}{\varepsilon} A'(x), \tag{48}$$

where $x$ is given in (28). Similarly,

$$W^*_{SF,11} = 2A'(x)\left(\frac{\varepsilon + (2-\varepsilon)\lambda_1^{2(\varepsilon-2)}}{\varepsilon}\right) + 4A''(x)\left(\lambda_1 - \lambda_1^{2\varepsilon-3}\right)^2, \tag{49}$$

and the bifurcation equation (47) becomes

$$\left(2A'\left(\varepsilon + (2-\varepsilon)\lambda_1^{2(\varepsilon-2)}\right) + 4\varepsilon A''\left(\lambda_1 - \lambda_1^{2\varepsilon-3}\right)^2\right)^{1/2} (A')^{1/2}\left(\lambda_1 - \lambda_1^{2\varepsilon-3}\right) +$$
$$2\lambda_1 A'\left(1 + (3-2\varepsilon)\lambda_1^{2(\varepsilon-2)}\right) + 4\lambda_1 A''\left(\lambda_1 - \lambda_1^{2\varepsilon-3}\right)^2 = 0. \tag{50}$$

There is an equivalent, more succinct form of this equation in terms of derivatives of $\tilde{W}$. It follows from (18) and (48) that

$$W^*_{SF,22} = \frac{2\tilde{W}'}{\varepsilon\left(\lambda_1 - \lambda_1^{2\varepsilon-3}\right)}. \tag{51}$$

Substituting both this and (18), (20) into (47) yields

$$\left( (\varepsilon - 1)^2 \lambda_1^{2(\varepsilon-2)} + \frac{\varepsilon(\lambda_1 - \lambda_1^{2\varepsilon-3})\widetilde{W}\,''}{2\widetilde{W}\,'} \right)^{1/2} \widetilde{W}\,' + \lambda_1 \widetilde{W}\,'' = 0, \tag{52}$$

where

$$\widetilde{W} = \widetilde{W}\left( \lambda_1^2 + \lambda_1^{2(\varepsilon-1)} - 2 + \frac{\varepsilon}{1-\varepsilon}\left( \lambda_1^{2(\varepsilon-1)} - 1 \right) \right). \tag{53}$$

Simplifications occur for the limiting values, $\varepsilon = 0, 1$. Letting $\varepsilon \to 0$ in (52) recovers the bifurcation criterion for perfectly incompressible materials

$$\lambda_1^3 \widetilde{W}\,'' + \widetilde{W}\,' = 0, \tag{54}$$

where

$$\widetilde{W} = \widetilde{W}\left( \lambda_1^2 + \lambda_1^{-2} - 2 \right). \tag{55}$$

This equation was first obtained by Reddy [26] (see also Dowaikh and Ogden [8]). This recovery further supports the choice of the kinematic relationship (9) between the pre-stretches assumed here.

Letting $\varepsilon \to 1$ yields

$$\left( (\lambda_1^2 - 1)\widetilde{W}\,' \right)^{1/2} + \left( 2\lambda_1^3 \widetilde{W}\,'' \right)^{1/2} = 0, \widetilde{W}\,'' \neq 0, \tag{56}$$

where

$$\widetilde{W} = \widetilde{W}\left( \lambda_1^2 - 1 - 2\ln \lambda_1 \right). \tag{57}$$

For the compressible neo-Hookean material (35),

$$\widetilde{W} = \frac{\mu}{2}\left( \lambda_1^2 + \lambda_1^{2(\varepsilon-1)} - 2 + \frac{\varepsilon}{1-\varepsilon}\left( \lambda_1^{2(\varepsilon-1)} - 1 \right) \right),$$
$$\widetilde{W}\,' = \mu(\lambda_1 - \lambda_1^{2\varepsilon-3}), \widetilde{W}\,'' = \mu\left( 1 + (3 - 2\varepsilon)\lambda_1^{2(\varepsilon-2)} \right). \tag{58}$$

The bifurcation equations (52), (54), (56) then become respectively

$$\left( \frac{\varepsilon + (2-\varepsilon)\lambda_1^{2(\varepsilon-2)}}{2} \right)^{1/2} \left( \lambda_1 - \lambda_1^{2\varepsilon-3} \right) + \lambda_1\left( 1 + (3 - 2\varepsilon)\lambda_1^{2(\varepsilon-2)} \right) = 0,$$
$$\lambda_1^6 + \lambda_1^4 + 3\lambda_1^2 - 1 = 0,$$
$$\lambda_1^2 - 1 + \lambda_1\left( 2(\lambda_1^2 + 1) \right)^{1/2} = 0. \tag{59}$$

The second of these is found in Biot [3] or Nowinski [27,28] and its unique positive root, denoted by $\lambda_c$, is given by $\lambda_c = 0.544$. The unique positive root of (59)$_3$ is given by

$$\lambda_c = \sqrt{-2 + \sqrt{5}} = 0.486. \tag{60}$$

Figure 1 displays a plot of the critical stretches versus the compressibility parameter $\varepsilon$.

FIGURE 1

Figure 1. Critical stretch values for surface stability versus the compressibility parameter ε, for a compressible neo-Hookean solid. When ε = 0, the critical stretch is 0.544, the value found for an incompressible neo-Hookean material. As ε increases, but remains small, the effect of finite compressibility is to decrease slightly the value of the critical compression stretch.

It is clear from this figure that the compressibility has little effect on the critical stretch for the compressible neo-Hookean material (35), with the incompressible value of $\lambda_c = 0.544$ a good approximation for realistic values of $\varepsilon$. In particular, this plot shows that the real slight compressibility of solid rubbers does *not* have an important effect on the critical stretch ratio of compression, at least for the material considered here. It simply makes the neo-Hookean semi-infinite solid slightly more stable in plane strain compression, by allowing it to be compressed by a few more percent before the critical stretch is reached. Therefore, slight compressibility does not provide an explanation as to why rubber blocks buckle in compression at ratios much higher than 0.544 (see [15]).

## 7. Calculation for the wave speed

The general wave speed equation is the secular equation (44). For the example of the general strain-energy function (26), this wave speed equation can be expressed simply in terms of derivatives of $\breve{W}$ as follows, using (18), (20) and (51):

$$\left(\varepsilon \lambda_1^2 \breve{W}'' + \frac{2(\varepsilon-1)^2}{\lambda_1^{3-2\varepsilon} - \lambda_1^{-1}} \breve{W}' - \varepsilon X\right)^{1/2} (\lambda_1 \breve{W}' - X) +$$

$$\sqrt{2}\left(\frac{\lambda_1^3 \breve{W}'}{\lambda_1^2 - \lambda_1^{2(\varepsilon-1)}} - X\right)^{1/2} (\lambda_1^2 \breve{W}'' - X) = 0, \tag{61}$$

where the notation of (41) has been used, $\breve{W} \equiv \tilde{W}/\mu$, and $\tilde{W}$ is given by (53). Thus the wave speed only depends on the strain-energy function through the derivatives of $\breve{W}$. Noting (45), solutions are now sought to this equation in the subsonic range

$$0 < X \leq \min\left(\frac{\lambda_1^3 \breve{W}'}{\lambda_1^2 - \lambda_1^{2(\varepsilon-1)}}, \lambda_1^2 \breve{W}'' + \lambda_1^{2(\varepsilon-2)} \frac{2(\varepsilon-1)^2}{\varepsilon} \frac{\lambda_1^3 \breve{W}'}{\lambda_1^2 - \lambda_1^{2(\varepsilon-1)}}\right). \tag{62}$$

As for the bifurcation equation, the form of the wave speed equation for the limiting values of $\varepsilon$ is again considered here. Letting $\varepsilon \to 0$ yields

$$\left(\frac{\lambda_1 \breve{W}'}{\lambda_1^4 - 1}\right)^{1/2}\left(\lambda_1 \breve{W}' - X\right) + \left(\frac{\lambda_1^5 \breve{W}'}{\lambda_1^4 - 1} - X\right)^{1/2}\left(\lambda_1^2 \breve{W}'' - X\right) = 0, \tag{63}$$

where $\breve{W} \equiv \tilde{W}/\mu$ and $\tilde{W}$ is given by (55). This is precisely the equation for perfectly incompressible materials obtained by Dowaikh and Ogden [8], albeit in a different form. Assuming that

$$\frac{\lambda_1^3 \breve{W}'}{\lambda_1^2 - \lambda_1^{2(\varepsilon-1)}} > 0, \ \breve{W}'' > 0, \tag{64}$$

means that the restriction (62) on the wave speed as $\varepsilon \to 0$ reduces to

$$0 < X \leq \frac{\lambda_1^3 \breve{W}'}{\lambda_1^2 - \lambda_1^{-2}}. \tag{65}$$

As $\varepsilon \to 1$, the wave speed equation reduces to

$$\left(\lambda_1^2 \breve{W}'' - X\right)^{1/2}\left[\left(\lambda_1 \breve{W}' - X\right) + \sqrt{2}\left(\frac{\lambda_1^3 \breve{W}'}{\lambda_1^2 - \lambda_1^{-2}} - X\right)^{1/2}\left(\lambda_1^2 \breve{W}'' - X\right)^{1/2}\right] = 0, \tag{66}$$

where $\breve{W} \equiv \tilde{W}/\mu$, $\tilde{W}$ is given by (57) and

$$0 < X \leq \min\left(\frac{\lambda_1^3 \breve{W}'}{\lambda_1^2 - 1}, \lambda_1^2 \breve{W}''\right). \tag{67}$$

As an example, consider the compressible neo-Hookean material (35). The wave speed equation (61) then becomes

$$\left(\varepsilon\lambda_1^2 + \lambda_1^{2(\varepsilon-1)}(2-\varepsilon) - \varepsilon X\right)^{1/2}\left(\lambda_1^2 - \lambda_1^{2(\varepsilon-1)} - X\right) + \\ \sqrt{2}\left(\lambda_1^2 - X\right)^{1/2}\left(\lambda_1^2 + (3-2\varepsilon)\lambda_1^{2(\varepsilon-1)} - X\right) = 0, \tag{68}$$

where

$$0 < X \leq \lambda_1^2, \tag{69}$$

independent of $\varepsilon$.

Consider the limiting special cases first. Letting $\varepsilon \to 0$ in (68) yields

$$\lambda_1^2 - \lambda_1^{-2} - X + \lambda_1(\lambda_1^2 - X)^{1/2}(\lambda_1^2 + 3\lambda_1^{-2} - X) = 0, \tag{70}$$

the same equation, albeit in a different form, as that given in Dowaikh and Ogden [8].

Letting $\varepsilon \to 1$ in (68) yields

$$(\lambda_1^2 + 1 - X)^{1/2}(\lambda_1^2 - 1 - X + \sqrt{2}(\lambda_1^2 - X)^{1/2}(\lambda_1^2 + 1 - X)^{1/2}) = 0. \tag{71}$$

One solution is immediate: $X = 1 + \lambda_1^2$, which however violates (69). The other solutions are obtained from an obvious squaring process of the other branch of the equation and are given by

$$X = 2 + \lambda_1^2 \pm \sqrt{5}. \tag{72}$$

Thus the only solution consistent with (69) is given by

$$X = \lambda_1^2 + 2 - \sqrt{5}. \tag{73}$$

Figure 2 below displays the variation of the surface wave speed versus the stretch ratio $\lambda_1$:

FIGURE 2

Figure 2. Normalized surface wave speed dependence on the pre-stretch for compressible neo-Hookean materials.

For ease of comparison, the wave speed is normalized with respect to the bulk shear wave speed in the deformed solid. At $\lambda_1 = 1$, the solid is isotropic. As $\lambda_1$ increases (plane strain tension), the wave speed increases too and has the shear wave speed as an upper bond. As $\lambda_1$ decreases (plane strain compression), the surface wave speed decreases and drops to zero when $\lambda_1$ reaches the critical stretch ratio. The thick curves are the plots of the limiting behaviours $\varepsilon = 0$ and $\varepsilon = 1$. The thin curve corresponds to the case $\varepsilon = 0.5$. That plot is very close to that of the ideally incompressible case, showing that compressibility, at least for the compressible neo-Hookean material, has very little effect on the wave speed, beyond a tendency to slightly lower the speed when the solid is in tension.

**Figure1**

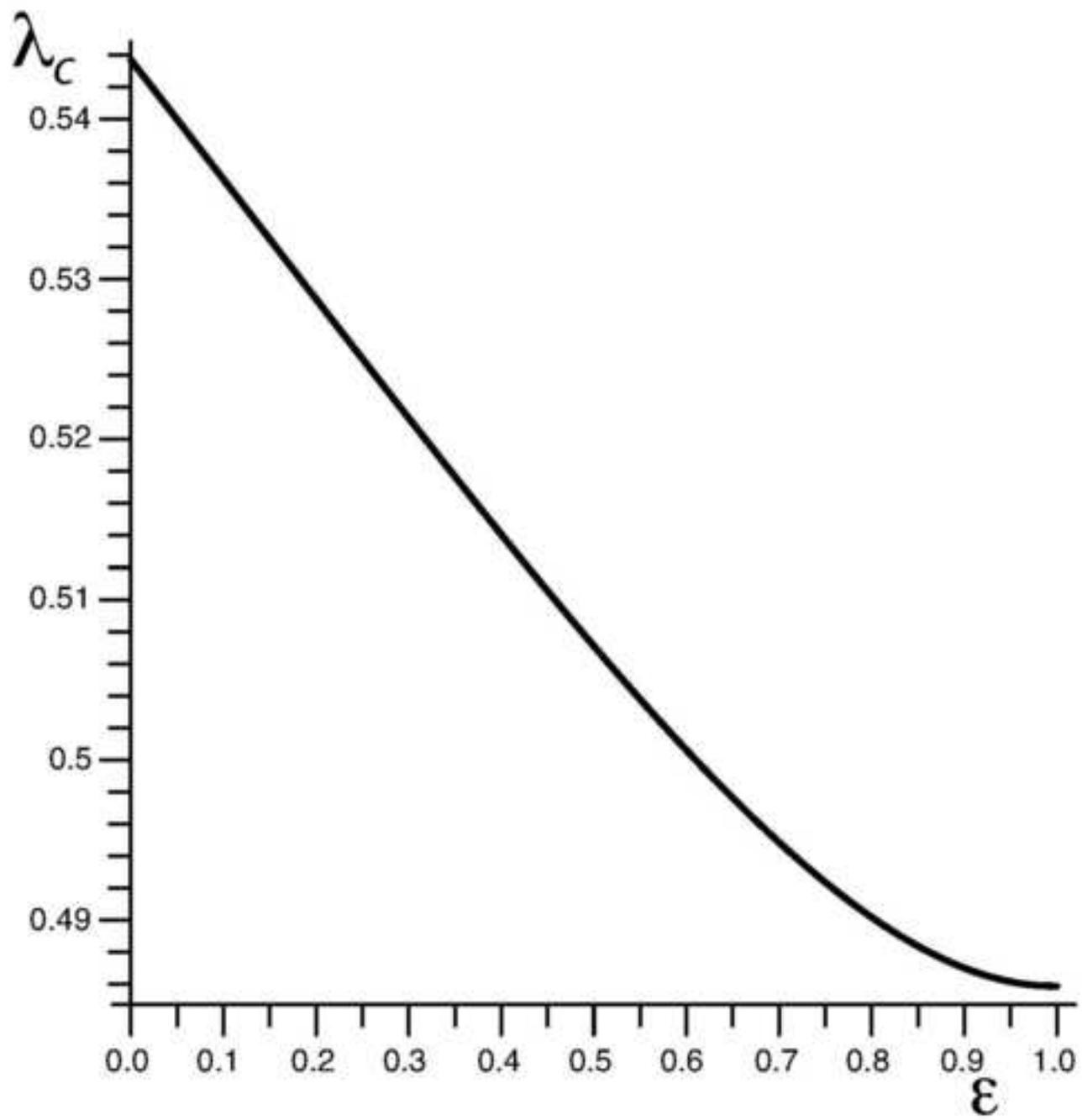

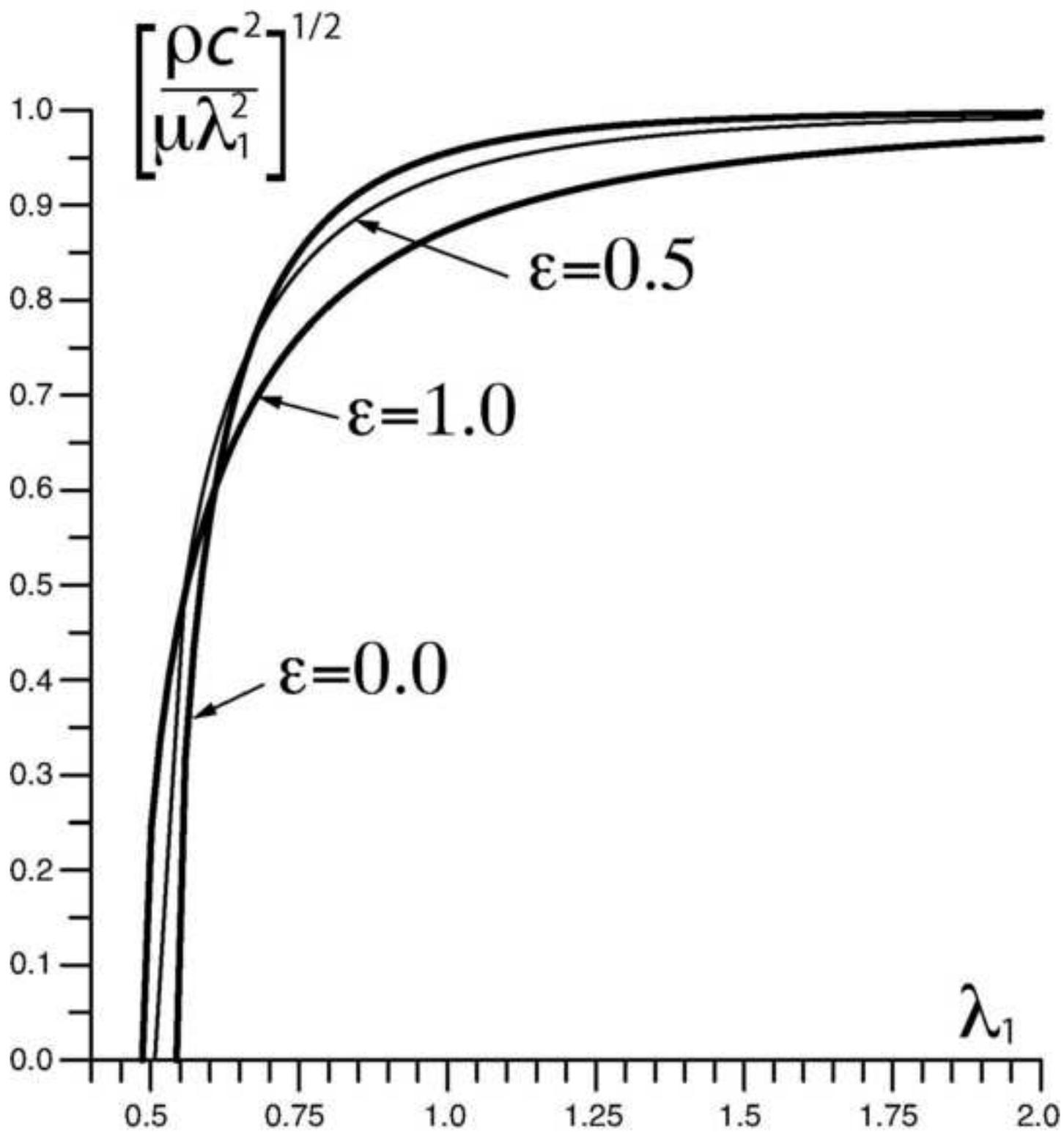